\documentclass[useAMS,usenatbib]{mn2e}
\title[Overshooting approach on Li abundance in clusters]{The diffusive overshooting approach on Li abundance in clusters}
\author[Q. S. Zhang]{Q. S. Zhang$^{1,2,3}$\thanks{E-mail: zqs@ynao.ac.cn (ZQS)}
\footnotemark[1]\\
$^{1}$National Astronomical Observatories/Yunnan Observatory, Chinese Academy of Sciences, P.O. Box 110, Kunming 650011, China.\\
$^{2}$Laboratory for the Structure and Evolution of Celestial Objects, CAS.\\
$^{3}$Graduate School of Chinese Academy of Sciences, Beijing 100039, China.}
\begin{document}

\date{Accepted . Received ; in original form }

\pagerange{\pageref{firstpage}--\pageref{lastpage}} \pubyear{2012}

\maketitle

\label{firstpage}

\begin{abstract}
The helioseimic investigation shows that the convective overshooting can penetrate $0.37H_P$ to the location where the temperature is $2.5\times10^6 K$ which is the typical temperature of the reaction $Li^7(p,\alpha)He^4$. This indicates that the overshooting mixing should be involved in investigating the solar Li abundance problem. Observations of Li abundance of solar-twins show that the sun is not very peculiar. The overshooting mixing should be also involved in investigating Li abundance in clusters. However, the fully overshooting mixing with the length of $0.37H_P$ results in too much Li depletion to fit the observation in the solar case. Therefore, using the diffusive process to describe the overshooting is more suitable. The diffusive overshooting approach requires the turbulent r.m.s. velocity in the overshooting region to calculate the diffusion coefficient. Turbulent convection models(TCMs), which are suggested by helioseimic investigation, can provide the turbulent properties in the overshooting region. However, TCMs are often too complex to be applied in the calculations of the stellar evolution. It is an easier way to use the asymptotic solution of TCMs. In this paper, I use the asymptotic solution of Li \& Yang's TCM, which results in agreements in both solar sound speed and solar Li abundance, to investigate Li abundance in clusters(Hyades, Praesepe, NGC6633, NGC752, NGC3680 \& M67). It is found that the overshooting mixing leads to significant Li depletion in the old clusters($t>1Gyr$) and little effect in the young clusters($t<1Gyr$).
\end{abstract}


\section{Introduction}

Li is a fragile element which can be burned via the reaction $Li^7(p,\alpha)He^4$ at the typical temperature $T\approx2.5\times10^6K$. Therefore, the abundance of Li at the stellar surface is a good tool to probe the property of the mixing in the stellar interior. Li abundance can be derived by the spectroscopic analysis. Abundant observations of Li abundance are accumulated in the recent few decades.

The observations of open clusters show some general relations between Li abundance and $T_{eff}$(see e.g., \citet{xio09} \textbf{and references therein}). There are three remarkable properties: i) Li abundance decreases when $T_{eff}$ decreases, ii) there is a Li gap near $T_{eff}\approx6500K$ in the clusters with the age no younger than the Hyades's, iii) the observations show dispersions of Li abundance which can't be thought as the observational uncertainties. The second could be due to the atomic diffusion and radiative acceleration(i.e., \citet{mi86,mi91,xio09}).

There are some inconsistencies between the observations and the relation of Li abundance vs. $T_{eff}$ predicted by the standard stellar evolution model. With the mixing length parameter of solar calibration and the old solar chemical composition($Z \approx 0.02$, \citet{gn93,gs98}) which results in correct location of the base of the solar convection zone, the standard stellar evolution model leads to too much Li depletion during the pre-main sequence(PMS) stage to fit the observations of young open clusters\citep{ve98b,pt02}. On the other hand, based on the standard stellar evolution model, which uses the convection as the only process of mixing, it has been found that Li depletion almost doesn't occur during main sequence(MS) stage for solar mass stars. This fails to explain the observations especially for the old clusters in the low-temperature rang, and there should be the extra-mixing process to deplete Li during MS\citep{dm84,pm89,dm94,pt02,dm03}. \citet{ri02,ri05} have investigated low-metallicity stars and have found that extra turbulent mixing is required to offset the atomic diffusion.

There are some possible mechanisms to deplete Li in MS: mass-loss\citep{hob89,sch90,swe92}, rotational mixing\citep{pin90,pin99,char92,char94}, internal waves mixing\citep{mon94,mon96,mon00}, diffusion and turbulent mixing\citep{mi86,mi91,ri02,ri05}, and overshooting mixing\citep{str76,sch99,xio09}. \citet{sr05} have studied the evolution of Li abundance and have concluded that "gravity waves can be excluded as the main agent responsible for MS Li depletion, while slow mixing induced by rotation might explain to some extent the empirical behavior". The rotational mixing, which is the most popular extra-mixing model, might reproduce the observational properties of the cluster Li abundance not only on the depletion but also on the dispersion(e.g., \citet{pin99}). Recently, \citet{chr11} have found that the convective overshooting could penetrate $0.37H_P$($H_P$ is the local pressure scale height) to the location where $T=2.5\times10^6 K$. This indicates that the overshooting mixing should be involved in studying the solar Li problem. Investigations on Li abundance of solar-twins shows that the solar Li abundance seems to be not very different comparing with solar-twins(see e.g., \citet{ki97,mr07,lt10,cas11}). Therefore, the overshooting mixing should be also involved in investigating the Li abundance in open clusters. Using Xiong's(1985) theory of non-local convection, \citet{xio09} have found that Li is significantly depleted during the MS stage when the overshooting mixing is present.

Recent helioseismic investigations have suggested using turbulent convection models(TCMs) to deal with the overshooting\citep{chr11}. The TCMs are based on fully hydrodynamic moment equations\citep{xio85,xio01,can97,can98b,can99,den06,li07,yan07}. Therefore, they are more physically reasonable than the phenomenological theories in the framework of mixing length\citep{hin75}. However, the TCMs are highly non-linear and are hard to be applied in stellar structure and evolution. Based on the TCM\citep{li07,yan07}, \citet{zha12b} have investigated the overshooting region and have found the asymptotical solution of the turbulent fluctuations in the overshooting region. Using the diffusive overshooting approach with the asymptotical solution, \citet{zha12a} have found that the solar sound speed and the solar Li abundance can be in agreement with the observations.

In this paper, I use the diffusive overshooting approach and the asymptotical solution of the Li \& Yang's(2007) TCM to investigate the Li abundance in six open clusters: Hyades, Praesepe, NGC6633, NGC752, NGC3680 \& M67. The stellar modeling method is described in Section 2. The diffusive overshooting approach is introduced in Section 3. The numerical results are shown in Section 4. Main conclusions and discussions are in Section 5.

\section{Stellar modeling}

In order to investigating Li abundance in clusters, I calculate the evolution series of the stellar models with different mass, metallicity and age. Six clusters are taken into account: three intermediate ones(Hyades, Praesepe \& NGC6633) and three old ones(NGC752, M67 \& NGC3680).

The modified stellar evolution code\citep{paz69}, which has been originally described by Paczynski and Kozlowski and has been updated by Sienkiewicz, is used to calculate the stellar evolution models. The OPAL equation of state tables \citep{rog96} are used. The OPAL opacity tables \citep{igl96} are used in the region of $lgT>3.95$, and the Alexander's opacity tables \citep{ale94} are used in the region of $lgT<3.95$. The nuclear reaction rates are from \citet{bah95} in the calculations.

The metal mixture of all clusters is assumed as the solar metal mixture\citep{gs98}. The metallicity of each cluster is calculated as:
\begin{equation}
Z   = 10^{[Fe/H]}(\frac{Z}{X})_{\odot}X
\end{equation}%
where $(\frac{Z}{X})_{\odot}=0.023$ is the ratio of the metallicity to the hydrogen abundance at the solar surface\citep{gs98}, $[Fe/H]$ the metallicity dex of corresponding cluster, and $X=0.7$ the initial hydrogen abundance of all stellar models. The solar composition used in this paper is not the latest observation(e.g., \citet{ags09}), for the reason that the new observational composition results in inconsistency on the location of the base of the solar convection zone between the solar model and the helioseismic data\citep{bas04,bah05,yang07,chr09,bi11}. However, the location of the base of the convective envelope(BCE) is crucial for calculating the Li abundance, since Li burning speed is mainly determined by the temperature of the BCE. Therefore, the solar composition of \citet{gs98}, which leads to the correct convective boundary in the solar model(e.g., \citet{bah05,pa11}), is adopted.

\begin{table}
\begin{center}
\caption{The age and metallicity dex of clusters. The references of the age and Li abundance of each cluster are as follows: (1)\citet{hp86}; (2)\citet{so93}; (3)\citet{so93b}; (4)\citet{bal95}; (5)\citet{din95}; (6)\citet{je97}; (7)\citet{pe98}; (8)\citet{jo99}; (9)\citet{ra00}; (10)\citet{pa01}; (11)\citet{je02}; (12)\citet{se04}; (13)\citet{bar09}. }
\begin{tabular}{crrrr}
  \hline
  Cluster & Age(Gyr) & Age Ref. & [Fe/H] & Li Ref.
  \\\hline
  Pleiades & 0.07 & 2 & -0.03 & 3  \\
  Hyades   & 0.6 & 7 & +0.13 & 4  \\
  Praesepe & 0.6 & 2 & +0.03 & 4  \\
  NGC6633  & 0.6 & 11 & -0.10 & 6  \\
  NGC752   & 2   & 1 & +0.01 & 4,12\\
  M67      & 4   & 5 & +0.05 & 8\\
  NGC3680  & 2   & 9 & -0.17 & 13,10
  \\\hline
\end{tabular}
\end{center}
\end{table}

The metallicity dex, age and the sources of Li abundance observations of concerned clusters are listed in Table 1. The mass ranges of stellar models of each cluster are: $0.80\sim1.20M_{\odot}$ for Hyades, Praesepe, NGC752 and M67, $0.70\sim1.20M_{\odot}$ for NGC6633 and NGC3680. The mass-step is $\triangle M=0.05M_{\odot}$.

The convection zone is assumed to be fully mixed. The convective boundary is defined by the Schwarzschild criteria. The convective heat flux is calculated by using the mixing length theory with the parameter $\alpha=2.08$ according to the calibration of the solar model by using those input physics.

All of the stellar model series evolve from the PMS stage with the center temperature $T_C=10^6K$ to the age of corresponding cluster(Table 1). The time step is no more than $0.5$ percent of the age of corresponding cluster. All of the stellar models include more than 2000 mesh points.

\section{The calculation of Li abundance and the overshooting approach}

\subsection{The diffusive overshooting mixing approach}

The equation of the evolution of Li abundance in the stellar interior is as follows:
\begin{equation}
\frac{\partial Y}{\partial t} = \frac{\partial}{\partial M_r}[(4\pi r^2 \rho)^2D \frac{\partial Y}{\partial M_r}]-\rho R XY
\end{equation}%
where $Y$ is Li abundance fraction, $X$ the hydrogen abundance fraction, $R$ the ratio of the reaction $Li^7(p,\alpha)He^4$ calculated by \citet{cf88} and $D$ the diffusion coefficient due to the convection and the overshooting. The convection zone is assumed as fully mixed, thus $D=+\infty, (\nabla_R\geq\nabla_{ad})$, where $\nabla_R, \nabla_{ad}$ are the radiative temperature gradient and the adiabatic one.

Li abundance is calculated in two cases: i) no overshooting. ii) taking the downward overshooting of the BCE into account. If the overshooting is absent, $D=0,(\nabla_R<\nabla_{ad})$. If the overshooting is present, $D=D_{OV},(\nabla_R<\nabla_{ad})$, where $D_{OV}$ is the diffusion coefficient of the overshooting mixing.

\begin{figure}
\vbox to2.2in{\rule{0pt}{7.5in}}
\includegraphics{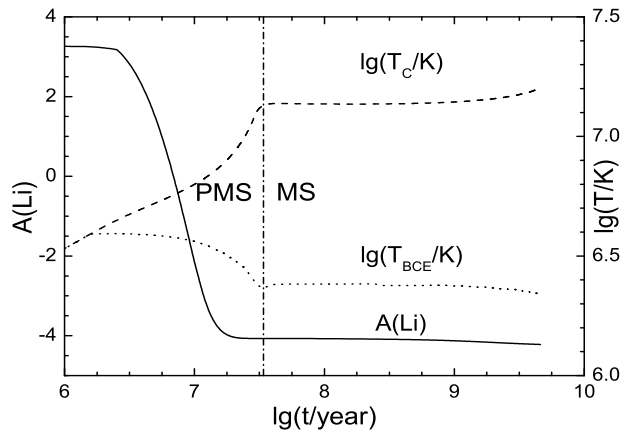}
 \caption{The evolutional history of surface Li abundance, center temperature $T_C$ and the BCE temperature $T_{BCE}$ of the solar model with fully overshooting mixing in $0.37H_P$.}
 \label{sample-figure}
\end{figure}

If the overshooting mixing is set to be instantaneous mixed as widely adopted in traditional, the sun shows excessive Li depletion. A fully mixed overshooting region with the length of $0.37H_P$\citep{chr11} results in $A(Li)\approx-4$ as show in Fig.1. It is too low to fit the observation $A(Li)_{\odot}=1.05$\citep{ags09}.

The fully overshooting mixing is excessively strong. The weak overshooting mixing described as a diffusive process is more suitable. The diffusion coefficient of overshooting mixing $D_{OV}$ is assumed as follows\citep{zha12a}:
\begin{equation}
D_{OV} = C_X H_P \sqrt{k}
\end{equation}%
where $C_X$ is a parameter and $C_X\sim10^{-10}$ according to \citet{zha12a} that is in consistent with \citet{den96}(Eqs.(27) \& (29)), $k$ the turbulent kinetic energy in the overshooting region.

The turbulent kinetic energy $k$ can be obtained by solving TCMs which are suggested by the helioseismic investigation\citep{chr11}. TCMs show that $k$ exponentially deceases in the overshooting region\citep{xio85,xio89,den06,zha09,zha12b}. This is in agreement with the numerical simulations(e.g., \citet{fre96}). It is a good choice to solve TCMs numerically to get the profile of $k$ in the overshooting region, but that is too complex. An easy way is to use the asymptotic solution of TCMs in the overshooting region. In the previous work\citep{zha12b}, we have investigated the TCM developed by \citet{li07} and have obtained the asymptotic solution in overshooting region as follows:
\begin{equation}
k=k_C(\frac{P}{P_C})^\theta,(Pe>1)
\end{equation}%
and
\begin{equation}
k=0,(Pe<1)
\end{equation}%
In Eqs.(4) and (5), $Pe=\frac{ \rho C_P l \sqrt{k}}{\lambda}$ is P\'{e}clet number, $P_C$ the pressure at the BCE, $\theta$ the exponential decreasing index of the turbulent kinetic energy $k$, and $k_C$ the turbulent kinetic energy at the BCE.

\begin{figure}
\vbox to2.2in{\rule{0pt}{7.5in}}
\includegraphics{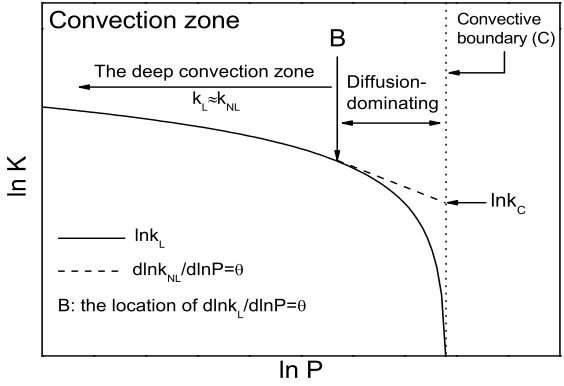}
 \caption{A sketch helps to understand the estimate of $k_C$ in non-local convection theory. Point 'B' is the location where the turbulent diffusion becomes significant. $k_L$ is turbulent kinetic energy determined by the localized TCM, and $k_{NL}$ is turbulent kinetic energy determined by the non-local TCM. }
 \label{sample-figure}
\end{figure}

It is crucial to get the turbulent kinetic energy at convective boundary(i.e., $k_C$) which appears in Eq.(4). A local convection theory(e.g., the mixing length theory) leads to $k=0$ at the convection boundary. In the TCM, the non-local effect(turbulent diffusion of $k$) is taken into account. In the framework of TCM of a convective zone it is possible to define a point 'B' separating two regions shown in Fig.2: the first one where the kinetic energy profile is well described by the local TCM: $k_{NL} \approx k_{L}$ (with $k_{NL}$ and $k_{L}$ the non-local and local kinetic energy respectively); and the second one between the point 'B' and the boundary of the convective region 'C' where the non local terms dominate and  $d\ln k_{NL}/d\ln P=\theta$ (with $\theta$ a constant determined only by the parameters of the TCM). As a consequence, the logarithmic value of the kinetic energy at the boundary of the convective zone ($ln k_C$) can be obtained by linear extrapolation from the value of $ln k_L$ at point 'B' with a slope $\theta$. The profile of $lnk_{NL}$ in the convection zone is the combination of the dashed line and the left part of the solid line to point B. Point 'B' is the the location where the turbulent diffusion becomes significant, so that $d\ln k_{L}/d\ln P=\theta$ at 'B'. This property results in the maximum of $k_C$(\textsl{the maximum of diffusion}, \citet{zha12b}).

According to \textsl{the maximum of diffusion}, $k_C$ can be estimated as follows\citep{zha12b}:
\begin{equation}
k_C^{\frac{3}{2}}=\frac{1}{e}k_B^{\frac{3}{2}}=\frac{1}{e}[\frac{ \alpha_1\beta g\lambda(\nabla_{R}-\nabla_{ad})}{\rho c_{P}}]_B
\end{equation}%
where $\beta=-(\partial ln \rho/\partial ln T)_P$ is the expansion coefficient, $g$ the gravity acceleration, $\lambda=(4acT^3)/(3\kappa\rho)$ the thermal conduction coefficient and the location of point 'B' can be calculated as follows\citep{zha12b}:
\begin{equation}
\sqrt{\frac{3}{4C_s \omega_C}} |\frac{r_C-r_B}{l}|=1
\end{equation}%
where $r_B$ and $r_C$ are the radius of B and the convective boundary respectively, $l=\alpha_1 H_P$, $\alpha_1,C_s,C_k$ are parameters of the TCM and $\omega_C=1/(3C_k)+1/3$.

The parameters of the TCM can be derived by experiments(e.g., \citet{gb76}), analyzing TCM itself (e.g., \citet{can98}) and the calibration of the solar model\citep{zha12a}. The appropriate parameter set of the TCM for the solar case is listed in \citet{zha12a}. Accordingly, the value of the parameters involved in this paper are: $\theta=-4,\alpha_1=0.93,C_s=0.08,C_k=2.5$.

Equation (4) is similar to Eq.(4) in Ventura et al.'s(1998) paper, since both of them describe the kinetic energy as a power law of the pressure in the overshooting region. The difference between Ventura et al.'s(1998) description and this paper is the calculation of the initial turbulent velocity. In Ventura et al.'s(1998) paper, it is determined by extrapolating the turbulent velocity distribution function resulting from the local convection theory near the convective boundary. In this paper, it is calculated based on the TCM which is a non-local convection theory.

\subsection{The initial Li abundance at ZAMS}

In the PMS stage, the stellar activity is significant thus the magnetic field could affect the stellar structure and the depletion of Li. Abundant observations show that the standard stellar theory underestimate the radii and overestimate the effective temperature of low mass stars \citep{to02,be06,mo08,mo10}. \citet{to06} and \citet{ri06} have suggested that those inconsistencies may result from the stellar activity because the surface activity and the magnetic field reduce the efficiency of convection and then the star should increase the radius to produce the required total radiative flux. With the mixing length parameter resulting from the solar calibration and the old solar chemical composition(e.g., \citet{gn93,gs98}), it has been found that\citep{ve98b,pt02} the standard stellar evolution model results in too high Li depletion during the PMS stage. That is also found in my calculations. It is shown in Fig.3 that the ZAMS Li abundance profile of Pleiades predicted by standard stellar model(the right dashed line) is too low to fit the observations. Applying the convection criterion modified by the magnetic field\citep{gt66,mo68}, \citet{ve98b} have found that the depletion of Li is lower than the value predicted by standard model and the difference is determined by the magnetic field intensity. On the other hand, the stellar activity has been thought to lead to the scatter of Li abundance in young clusters\citep{xio05,ki10,pin10}. \citet{xio05} have investigated the $\alpha$ Per cluster and have pointed out that the dispersion of Li abundance in the cluster can be caused by inhomogeneous reddening and stellar activity. \citet{ki10} have investigated the cool Pleiades dwarfs and have suggested that the striking Li dispersion is caused by surface activity inflating the star radii in the PMS stage. \citet{pin10} have showed that the Pleiades Li dispersion is reproduced when the radius of stellar models are inflated 10\%.

\begin{figure}
\vbox to2.2in{\rule{0pt}{7.5in}}
\includegraphics{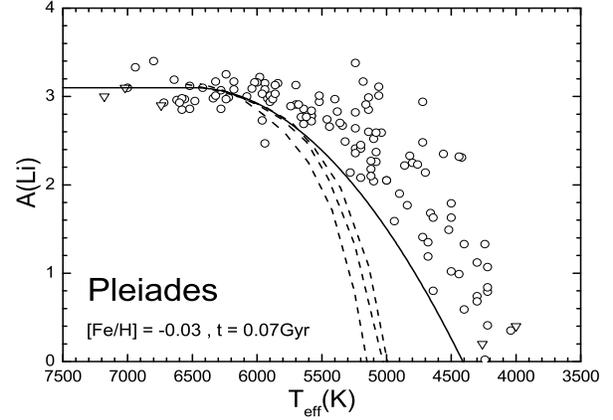}
 \caption{Li abundance of the young cluster Pleiades. Observation data: circles and triangles are the observational Li abundance and the upper limits respectively. The solid line is the profile of Eq.(8), and presents the approximate profile of the low envelop of the Pleiades Li abundance. Dashed lines are the ZAMS Li abundance predicted by standard stellar model in which the PMS Li depletion is taken into account. From left to right, the dashed lines correspond to the Hyades, Praesepe, Pleiades case, respectively. }
 \label{sample-figure}
\end{figure}

The effects of the stellar activity and the magnetic field on stellar structure are complex. The complexity makes it hard to trace the depletion of Li accurately during the PMS stage. This paper focuses on how the overshooting mixing affects Li depletion during MS stage. In order to avoid the problem of the PMS Li depletion affected by the stellar activity and the magnetic field, I use the relation of Li abundance vs. $T_{eff}$ of the young cluster Pleiades to be the initial Li abundance function at ZAMS for all clusters considered in this paper. The relation of Li abundance vs. $T_{eff}$ of Pleiades is shown in Fig.3. As mentioned above, the scatter of Li abundance could be due to the surface activity\citep{ki10,pin10}. Therefore, the lower envelope of Li abundances in Pleiades is set as the initial Li abundance of ZAMS stars. The solid line in Fig.3 is the function:
\begin{equation}
A(Li) = \left\{ {\begin{array}{*{20}{c}}
   {3.1 - 30{{(\frac{{{T_{eff}}}}{{6500K}} - 1)}^2};{T_{eff}} < 6500K}  \\
   {3.1;{T_{eff}} \ge 6500K}  \\
\end{array}} \right.
\end{equation}%
which is an approximate profile of the lower envelope of Li abundances. Equation (8) is used to calculate the Li abundance of ZAMS stars in this paper.

\section{Numerical results}

Using the stellar modeling method described in Section 2, I calculated stellar models with different mass in the case of six clusters(Hyades, Praesepe, NGC6633, NGC752, M67 \& NGC3680). Based on those stellar models, I calculated Li abundance in both cases with and without the overshooting by solving Eq.(2) with the initial Li abundance of ZAMS stars calculated via Eq.(8).

\subsection{Theoretical Li abundance isochrones with and without the overshooting mixing}

\begin{figure}
\vbox to 7.3in{\rule{0pt}{7.5in}}
\includegraphics{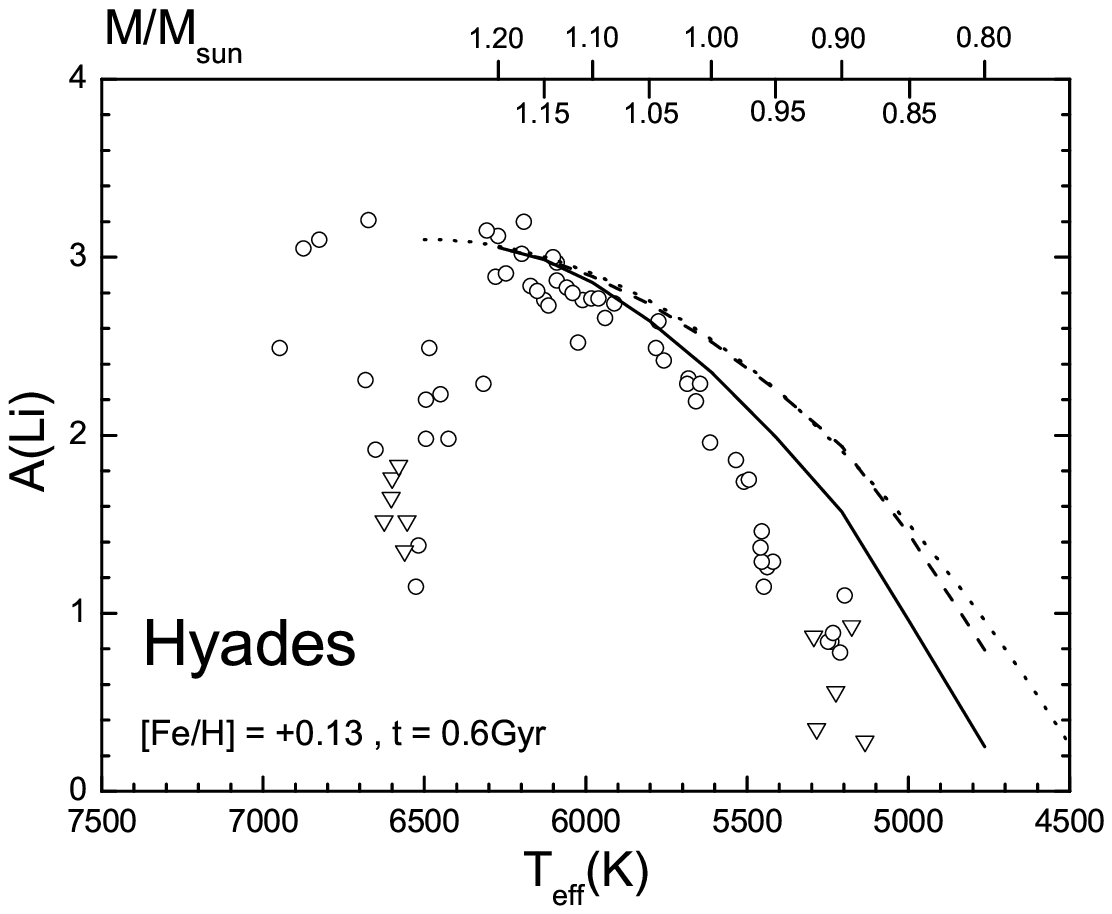}
\includegraphics{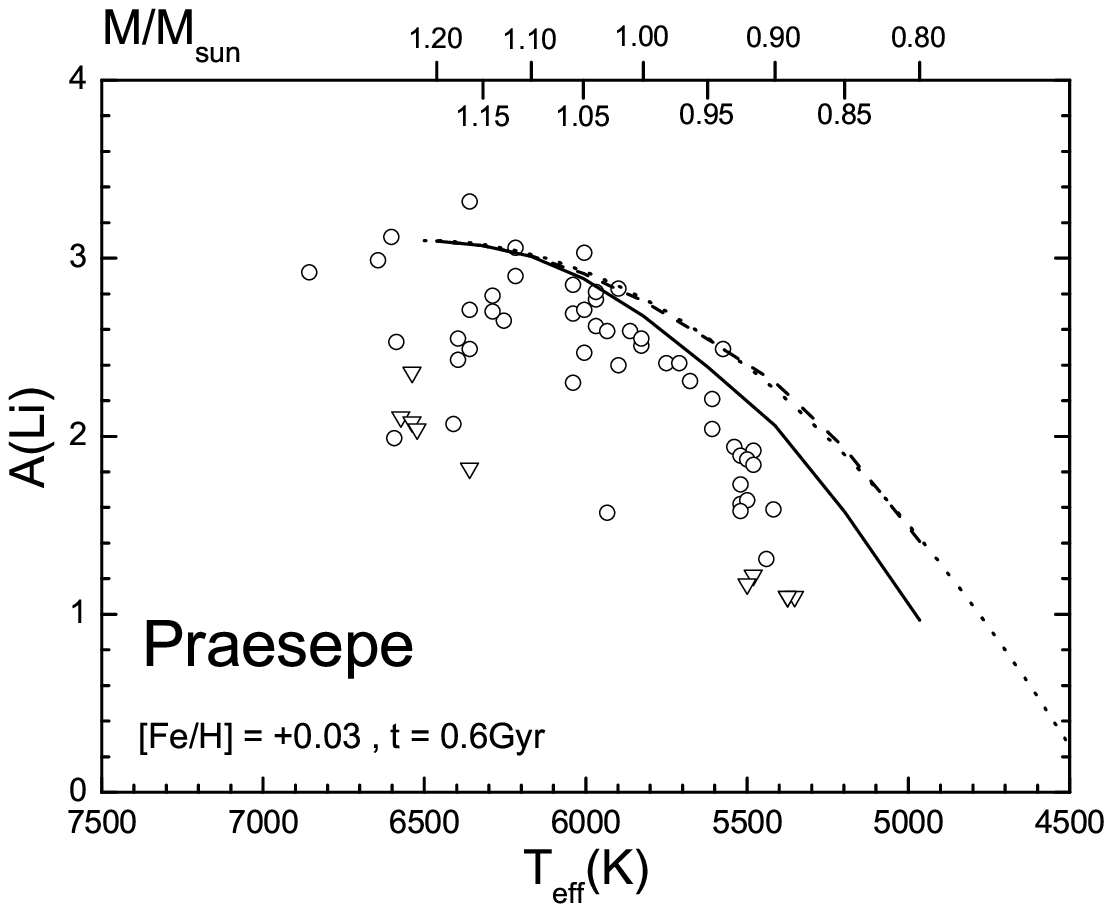}
\includegraphics{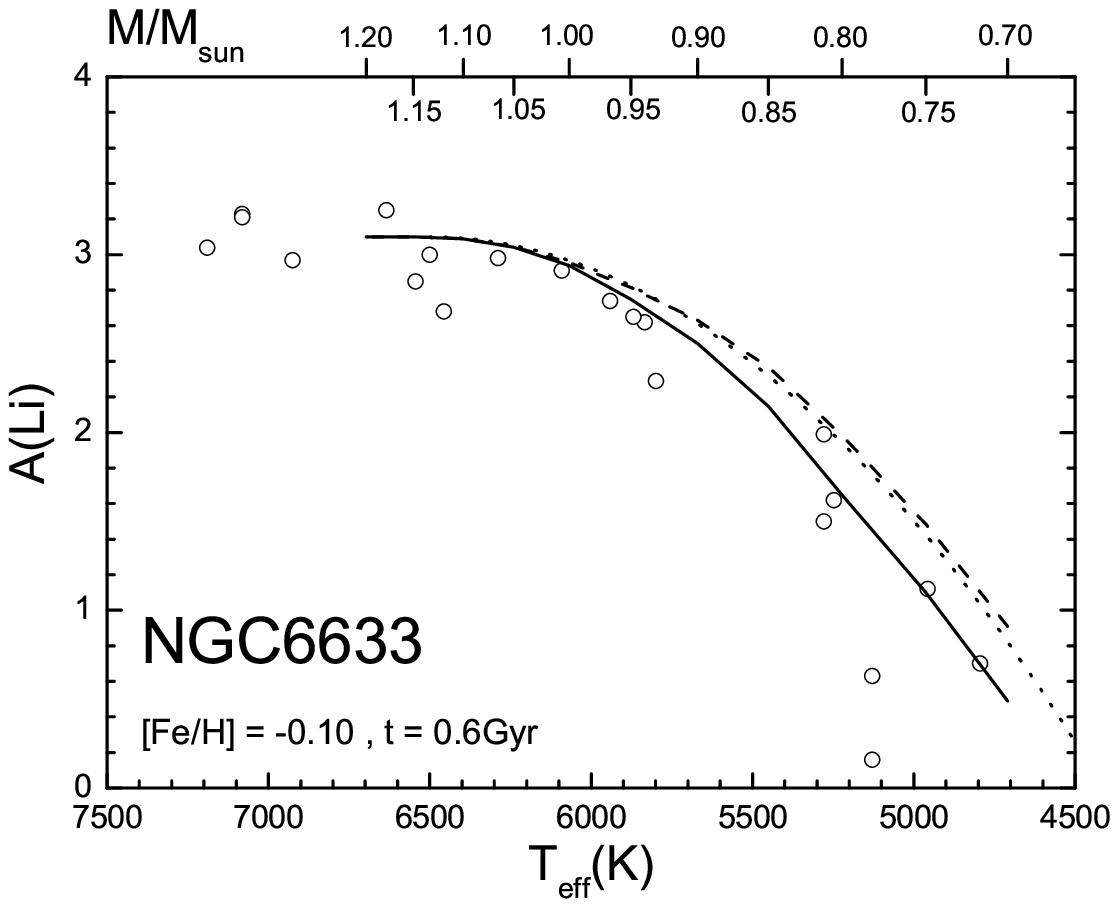}
 \caption{Li abundance of the intermediate clusters Hyades, Praesepe \& NGC6633. Observation data : circles and triangles are the observational Li abundance and the upper limits respectively. The dotted lines show the relation between Li abundance and $T_{eff}$ at ZAMS. The dashed lines show the Li abundance of stellar model without overshooting at corresponding $[Fe/H]$ and $t$, and the solid lines are with overshooting. The upper horizonal axis is the mass of stellar model, links to the dashed line and the solid one. The upper horizonal axis doesn't link to the dotted line. }
\label{sample-figure}
\end{figure}

The theoretical Li abundance isochrones of three intermediate clusters(Hyades, Praesepe \& NGC6633) are shown in Fig.4. The dotted-lines are the initial Li abundance according to Eq.(8). The dashed-lines are the theoretical Li abundance isochrones of stellar models without overshooting. The solid-lines are the theoretical Li abundance isochrones of the stellar models with overshooting.
It can be found that there is almost no Li depletion in those intermediate clusters in the concerned mass rang when the overshooting is absent. However, the overshooting case show modification on Li depletion. For lower $T_{eff}$ star, overshooting mixing results in more modification on Li depletion. Lower $T_{eff}$ represents higher temperature at the BCE, thus the overshooting mixing is more efficient on depleting Li. The overshooting mixing does deplete more Li, but the modifications seems to be not enough to explain the observations of clusters Hyades \& Praesepe. This is probably for the reason that the metallicity of Hyades and Praesepe is larger than Pleiades's and then Eq.(8) may overestimate the ZAMS Li abundance of Hyades and Praesepe. We can estimate how much Eq.(8) overestimates the ZAMS Li abundance of Hyades and Praesepe via Fig.3. It is found that the predicted ZAMS Li abundance of Pleiades is almost identical to Eq.(8) at the high temperature rang with $T_{eff}>5500K$. This indicates that the standard stellar model predicts ZAMS Li abundance well if $T_{eff}>5500K$. At $T_{eff}=5500K$, the ZAMS Li abundance profiles predicted by standard model show about 1.2, 0.9 and 0.8 dex depletion for Hyades, Praesepe and Pleiades, respectively. Accordingly, Eq.(8) overestimates about 0.4 dex of Li depletion for Hyades and 0.1 dex for Praesepe at $T_{eff}=5500K$. If these modifications are taken into account in the Hyades and Praesepe cases in Fig.4 at $T_{eff}=5500K$, the solid lines should be in better agreement with the observations. The gap of Li abundance in the rang $6200K<T_{eff}<6800K$ in the observations of the clusters(Hyades \& Praesepe) can't be reproduced by the theoretical models because of the absence of the possible mechanisms(e.g. atomic diffusion and radiative acceleration) in this work. It seems to be no Li gap in the cluster NGC6633, probably because the observation data in the corresponding rang of $T_{eff}$ are not enough to reveal the gap.

\begin{figure}
\vbox to 7.3in{\rule{0pt}{7.5in}}
\includegraphics{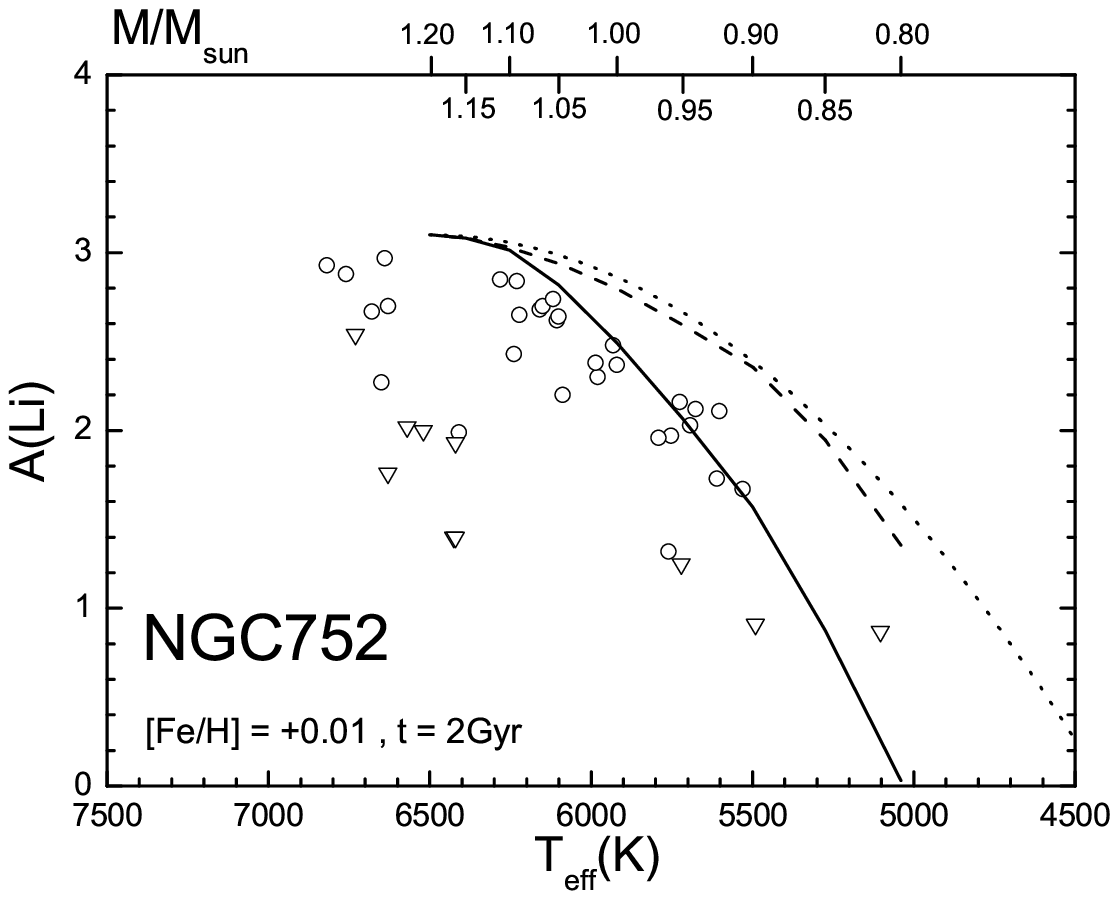}
\includegraphics{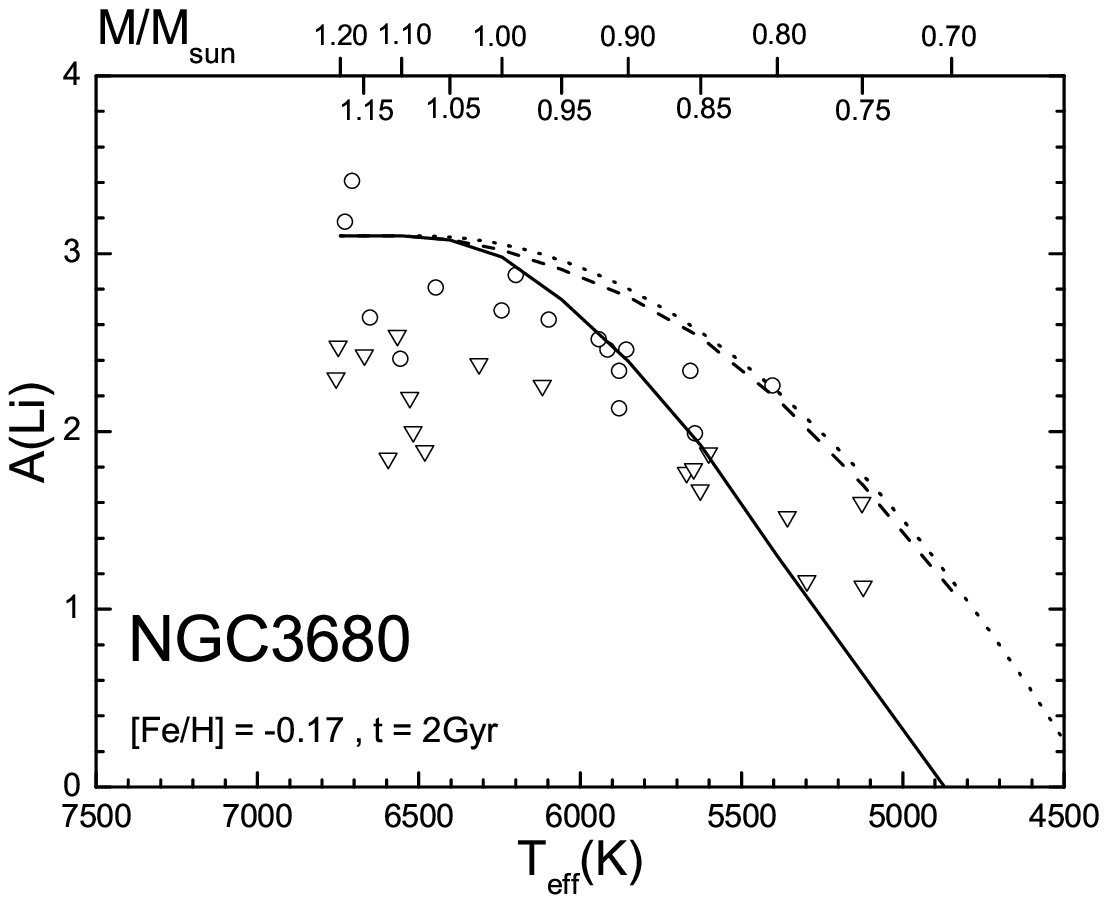}
\includegraphics{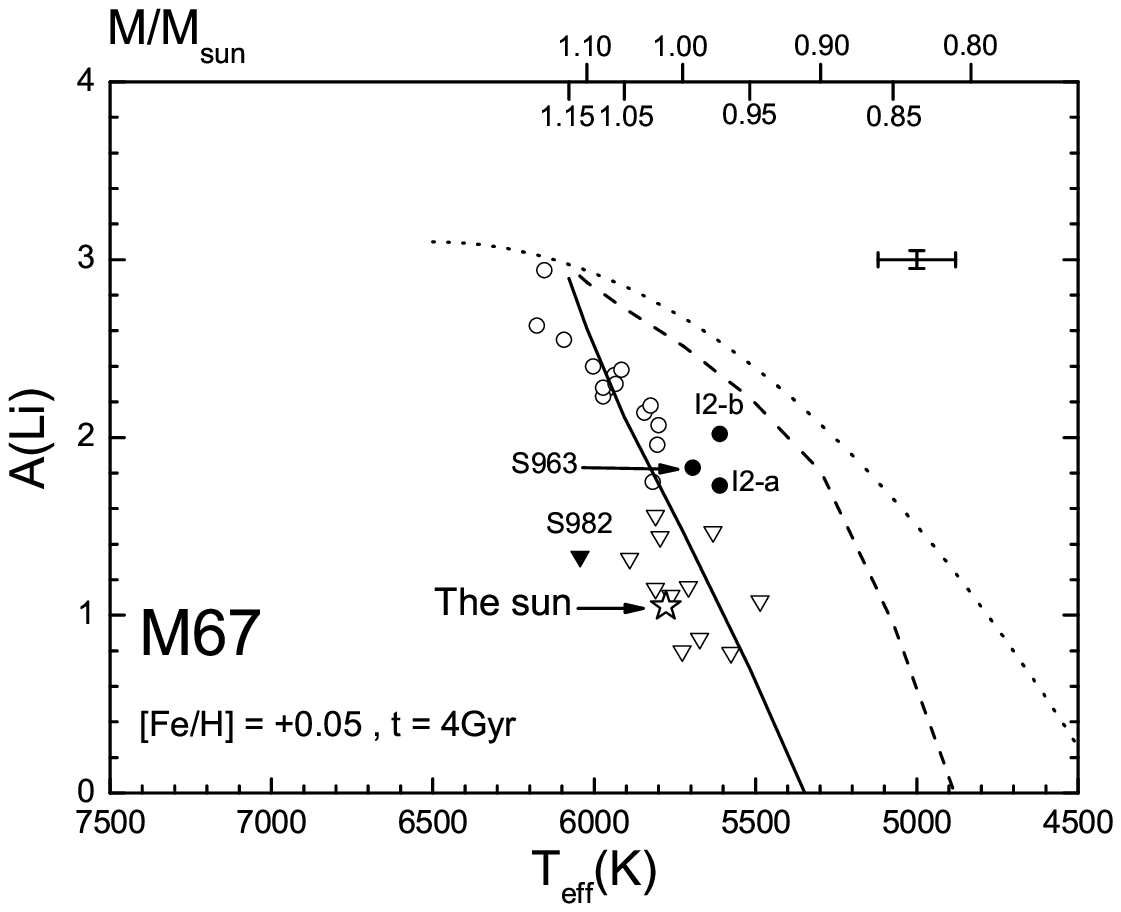}
 \caption{Similar to Fig.4, but for old clusters NGC752, NGC3680 \& M67. The solid symbols in the M67 case denote binary component. }
 \label{sample-figure}
\end{figure}

The theoretical Li abundance isochrones of three old clusters(NGC752, NGC3680 \& M67) are shown in Fig.5. The observations of Li abundance of NGC3680 and M67 include only the MS stars with the luminosity lower than the turn-off luminosity, in order to distinguish the evolved stars which may be in the same rang of $T_{eff}$ as the investigated stars. Li abundance of the sun is also denoted in the M67 case in Fig.5, since the age of the sun is close to M67's. The theoretical Li depletion of the stellar models without overshooting are obviously not enough since almost all the observation data locate below the theoretical isochrone. It is found that the overshooting mixing leads to significant modifications on Li abundance in the stellar models. This is because the age of those clusters are long, the overshooting mixing could deplete more Li comparing with the results of the intermediate clusters. The M67 cluster show a Li dispersion. It is probably because of the uncertainty of $T_{eff}$. In deriving the effective temperature of the samples of M67 in Jones et al.'s(1999) work, the uniform reddening value has been adopted. \citet{tay07} has studied the reddening of M67 cluster, and has found that the standard error of mean reddening value is $\sigma[\overline{E(B-V)}]=0.004mag$ by analyzing a sample including about 50 stars. Accordingly, the standard error of the reddening value for a star in M67 is $\sigma[E(B-V)]=0.03mag$(noting the statistical relation $\sigma(x)=\sqrt{N}\sigma(\overline{x})$). For the M67 stars investigated in this paper, 0.01 mag uncertainty of color corresponds to 40K uncertainty of $T_{eff}$(Jones et al., 1999). Therefore, the standard error of the reddening value for a M67 star leads to $\sigma (T_{eff})=120K$. It can be found in the M67 case in Fig.5 that the observations are in agreement with the Li vs. $T_{eff}$ relation predicted by the overshooting stellar models, when $\sigma (T_{eff})=120K$ is taken into account. There are some stars('I2-a', 'I2-b' and 'S982' in Fig.5) being outside $1\sigma$ range of $T_{eff}$, but they are binary components.

In this paper, the settling of Li is not included because the main purpose is to investigate how the overshooting affects the Li abundance of the G and K type low mass stars. The timescale of the settling is sensitive to the mass of star. It is very short for the massive star and is very long for the low-mass star. Therefore, the settling should not significantly affect the low mass stars. \citet{xio09} have found that the settling of Li is significant for $M>1.1M_{\odot}$ and is almost no effects for $M<1M_{\odot}$. Accordingly, if the settling is taken into account, the results of the low-mass star with $T_{eff}<5800K$ in this paper should not be changed significantly and the Li abundance of the stars with $T_{eff}>5800K$ should be depleted more.

\subsection{Li burning history and its depletion time scale}

\begin{figure}
\vbox to2.4in{\rule{0pt}{7.5in}}
\includegraphics{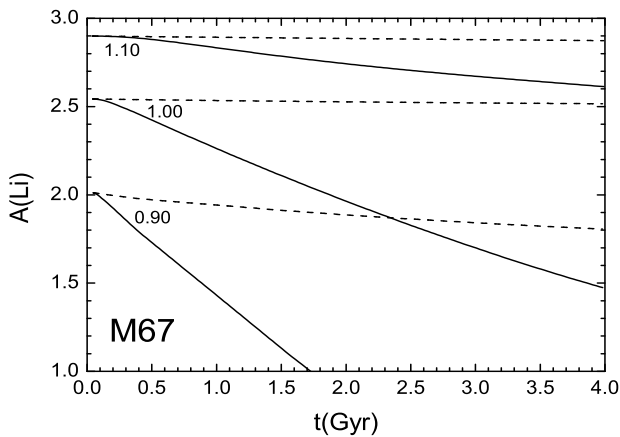}
 \caption{The evolution history of Li abundance of stellar models with $M=0.90M_{\odot},1.00M_{\odot},1.10M_{\odot}$ for the M67 case. Dash lines are for the stellar models without the overshooting and the solid lines are for the stellar models with the overshooting.}
 \label{sample-figure}
\end{figure}

As it is found in Figs.(4-5), the theoretical Li abundance isochrones of the stellar models with $M\geq 0.90M_{\odot}$ show only a little Li depletion during the MS stage when the overshooting is absent. In order to investigate the evolution of Li abundance of stellar models, the relation between $A(Li)$ and $t$ of stellar models with $M= 0.90, 1.00, 1.10 M_{\odot}$ for the old cluster M67 are shown in Fig.6. The solid lines show the Li abundance of stellar models with overshooting mixing, and the dashed lines correspond to the no overshooting case. For the overshooting case, it is found that the relation between $A(Li)$ and $t$ is almost linear. This indicates that the surface Li abundance satisfies the equation below in the overshooting case:
\begin{equation}
[Li]\approx[Li]_0 exp(-\frac{t-t_0}{\tau_{Li}})
\end{equation}%
where $\tau_{Li}$ is the time scale of Li depletion.

Now, I estimate $\tau_{Li}$ for the solar mass star. The time scale of Li depletion at radius $r$ in the star should be $\tau_{Li}(r)=Max(\tau_N(r),\tau_{OV}(r))$ where $\tau_N(r)$ is Li burning time scale at $r$ and $\tau_{OV}(r)$ is the overshooting mixing time scale. For solar mass star, $lgT_{BCE}\approx 6.35$ during the MS stage, and the typical temperature of the reaction $Li^7(p,\alpha)He^4$ at the location where $\tau_N(r)\sim1Gyr$ is about $lgT\approx 6.4$(denoting this location as A). Therefore, the mixing distance from the BCE to A is about $L\approx0.12H_T\approx0.3H_P$(noting that the temperature gradient $\nabla=dlnT / dlnP \approx\nabla_{ad}\approx0.4$ near the BCE) where $H_T=-dr/dlnT$. The time scale of the overshooting mixing process at A is $\tau_{OV}(A)\sim L^2 / D_{OV} \approx\ 0.3^2H_P/(C_X\sqrt{k})$ where $H_P\sim 10^{10}cm$ and $\sqrt{k}\sim 10^3cm/s$ near the BCE in the solar case, thus $\tau_{OV}(A)\sim10^{16}s\sim1Gyr$ and $\tau_N(A)\sim1Gyr$. $\tau_{OV}(r)>1Gyr$ in the region $lgT > 6.4$ and $\tau_{N}(r)>1Gyr$ in the region $lgT < 6.4$ because the reaction rate $R\propto T^{22}$. Therefore, the time scale of Li depletion of a solar mass star in the MS stage should be $\tau_{Li}=Min(\tau_{Li}(r))=\tau_{Li}(A)\sim1Gyr$.

It can be found in Fig.6 that $[Li]/[Li]_0 \approx 1/e$ at $t\approx1.5Gyr$ for the $1.00M_{\odot}$ star. This means $\tau_{Li}\approx 1.5Gyr$ for the $1.00M_{\odot}$ star and validates the estimate above.

\section{Conclusions and Discussions}

I have investigated Li abundance in six open clusters(Hyades, Praesepe, NGC6633, NGC752, NGC3680 \& M67) by using the diffusive overshooting approach, which describes the overshooting mixing as a diffusion process. The diffusion velocity is based on the turbulent convection model\citep{li07,zha12b}. The PMS Li depletion is affected by the stellar activity and the magnetic field. In order to take into account the PMS Li depletion and avoid the difficulty of stellar activity and the magnetic field, the relation between Li abundance and $T_{eff}$ of the young cluster Pleiades is set as the initial function of Li abundance vs. $T_{eff}$ for ZAMS stars. It is found that the diffusive overshooting mixing, which is based on the TCM with parameters favored by the sun, leads to remarkable Li depletion in MS stage for the old clusters.

There is almost no Li depletion for the stars with $M\geq0.9M_{\odot}$ during the MS stage when the overshooting is absent.
The time scale of Li depletion due to the overshooting mixing is about $\tau_{Li} \sim 1Gyr$ for the $1M_{\odot}$ star in the MS stage. For the young clusters($t<1Gyr$, e.g., Hyades, Praesepe, NGC6633), the modifications of the overshooting mixing on Li abundance are slight due to $t<\tau_{Li}$. However, for the old clusters($t>1Gyr$, e.g., NGC752, M67 \& NGC3680), the overshooting mixing should be included in order to fit the observations.

A main distinction between this work and Xiong \& Deng's(2009) work is whether the PMS depletion of Li is taken into account.
Xiong \& Deng's(2009) results show that Li is depleted to $A(Li)=0$ at $T_{eff}\approx5200K$ for Hyades and Praesepe, at $T_{eff}\approx5500K$ for NGC752, and at $T_{eff}\approx5700K$ for M67. Comparing Figs.(4) and (5) with Xiong \& Deng's(2009) results, which ignore the PMS Li depletion, the results in this paper show less Li depletion. These indicate that the overshooting mixing effects in this paper is much weaker than theirs. This is probably because of the small parameter(e.g., diffusion parameter $C_X$) in this model. The small parameter is due to the unknown characteristic scale of the overshooting mixing process, which should be in the range between the Kolmogorov scale and the largest scale in the convection zone, and the assumption that the time scale of the overshooting mixing should be comparable with the stellar evolutionary time scale\citep{den96}. Our previous work\citep{zha12a} showed that this weak mixing seems to be in consistent with the helioseismic data.

Noting that the overshooting mixing with the parameters favored by the sun is also suitable for Li abundance of low mass stars in old clusters, it is thought to be reasonable that the parameters of the TCM in the solar case\citep{zha12a} and the diffusion parameter $C_X \sim 10^{-10}$\citep{den96,zha12a} is generally applicable on the downward overshooting of the convective envelop of low mass stars.

On the Li abundance in clusters, the overshooting mixing is not the only candidate. There are other mechanisms can deplete Li as mentioned in Section 1. However, the helioseismic research\citep{chr11} have showed that the overshooting of the heat flux could penetrate $0.37H_P$ to the location where $T=2.5\times10^6 K$ in the solar case. And according to simulations\citep{si95,mk07,mk10} and TCMs\citep{xio01,zha12b}, the overshooting of turbulent kinetic energy, which represents the turbulent r.m.s. velocity and affects the mixing directly, can penetrate deeper than heat flux. This indicates that the overshooting mixing seems to be unavoidable in investigating Li abundance.

\section*{Acknowledgments}

 Many thanks to the anonymous referee for his/her productive and valuable suggestions. Fruitful discussions with Y. Li are highly appreciated. This work is co-sponsored by the National Natural Science Foundation of China through grant No.10673030 and No.10973035, Science Foundation of Yunnan Observatory No.Y1ZX011007, Yunnan Province under Grant No.2010CD112 and Chinese Academy of Sciences under grant no. KJCX2-YW-T24.

\label{lastpage}

\begin{thebibliography}{}
\bibitem [\protect\citeauthoryear{Alexander \& Ferguson}{1994}] {ale94} Alexander D.R., \& Ferguson J.W., 1994, ApJ, 437, 879
\bibitem [\protect\citeauthoryear{Asplund et al.}{2009}] {ags09} Asplund M., Grevesse N., Sauval A. J., \& Scott P. 2009, ARA\&A, 47, 481
\bibitem [\protect\citeauthoryear{Bahcall et al.}{1995}] {bah95} Bahcall J.N., Pinsonneault M.H., \& Wasserburg G.J., 1995, Rev. Mod. Phys., 67, 781
\bibitem [\protect\citeauthoryear{Bahcall et al.}{2005}] {bah05} Bahcall J. N., Basu S., Pinsonneault M., \& Serenelli A. M., 2005, ApJ, 618, 1049
\bibitem [\protect\citeauthoryear{Balachandran}{1995}] {bal95} Balachandran S., 1995, ApJ, 446, 203
\bibitem [\protect\citeauthoryear{Barbara et al.}{2009}] {bar09} Barbara J. et al. 2009, AJ, 138, 1171
\bibitem [\protect\citeauthoryear{Basu \& Antia}{2004}] {bas04} Basu S., \& Antia H. M., 2004, ApJ, 606, L85
\bibitem [\protect\citeauthoryear{Berger et al.}{2006}] {be06} Berger D.H., Gies D.R., McAlister H.A., et al. 2006, ApJ, 644, 475
\bibitem [\protect\citeauthoryear{Bi et al.}{2011}] {bi11} Bi S. L., Li T. D., Li L. H., \& Yang W. M., 2011, ApJ, 731, L41
\bibitem [\protect\citeauthoryear{Canuto}{1997}] {can97} Canuto V. M., 1997, ApJ, 482, 827
\bibitem [\protect\citeauthoryear{Canuto}{1998}] {can98b} Canuto V. M., 1998, ApJ, 508, 767
\bibitem [\protect\citeauthoryear{Canuto \& Dubovikov}{1998}] {can98} Canuto V. M., \& Dubovikov M., 1998, ApJ, 493, 834
\bibitem [\protect\citeauthoryear{Canuto}{1999}] {can99} Canuto V. M., 1999, ApJ, 524, 311
\bibitem [\protect\citeauthoryear{Castro et al.}{2011}] {cas11} Castro M. et al., 2011, A\&A, 526, A17
\bibitem [\protect\citeauthoryear{Caughlan \& Fowler}{1988}] {cf88} Caughlan Georgeann R.; Fowler William A., 1988, Atomic Data and Nuclear Data Tables, 40, 283.
\bibitem [\protect\citeauthoryear{Charbonnel et al.}{1992}] {char92} Charbonnel C., Vauclair S., \& Zahn J.-P., 1992, A\&A, 255, 191
\bibitem [\protect\citeauthoryear{Charbonnel et al.}{1994}] {char94} Charbonnel C., Vauclair S., Maeder A., Meynet G., \& Schaller G., 1994, A\&A, 283, 155
\bibitem [\protect\citeauthoryear{Christensen-Dalsgaard et al.}{2009}] {chr09} Christensen-Dalsgaard J., Di-Mauro M.P., Houdek G., \& Pijpers F., 2009, A\&A, 494, 205
\bibitem [\protect\citeauthoryear{Christensen-Dalsgaard et al.}{2011}] {chr11} Christensen-Dalsgaard J., Monteiro M.J.P.F.G., Rempel M., \& Thompson M.J., 2011, MNRAS, 414, 1158
\bibitem [\protect\citeauthoryear{D'Antona \& Mazzitelli}{1984}] {dm84} D'Antona F., \& Mazzitelli I., 1984, A\&A, 138, 431
\bibitem [\protect\citeauthoryear{D'Antona \& Mazzitelli}{1994}] {dm94} D'Antona F., \& Mazzitelli I., 1994, ApJS, 90, 467
\bibitem [\protect\citeauthoryear{D'Antona \& Montalb\'{a}n}{2003}] {dm03} D'Antona F., \& Montalb\'{a}n J., 2003, A\&A, 412, 213
\bibitem [\protect\citeauthoryear{Deng et al.}{1996}] {den96} Deng L., Bressan A., Chiosi C., 1996, A\&A, 313, 145
\bibitem [\protect\citeauthoryear{Deng et al.}{2006}] {den06} Deng L., Xiong D.R., \& Chan K.L., 2006, ApJ, 643, 426
\bibitem [\protect\citeauthoryear{Dinescu et al.}{1995}] {din95} Dinescu D.I., Demarque P., Guenther D.B., \& Pinsonneault M.H., 1995, AJ, 109, 2090
\bibitem [\protect\citeauthoryear{Freytag et al.}{1996}] {fre96} Freytag B., Ludwig H.-G., \& Steffen M., 1996, A\&A, 313, 497
\bibitem [\protect\citeauthoryear{Gibson \& Launder}{1976}] {gb76} Gibson M.M., \& Launder B.E., 1976, Trans. ASME, J. Heat Transfer, 98, 81
\bibitem [\protect\citeauthoryear{Gough \& Tayler}{1966}] {gt66} Gough D.O.,\& Tayler R.J. 1966, MNRAS, 133, 85
\bibitem [\protect\citeauthoryear{Grevesse \& Noels}{1993}] {gn93} Grevesse N., \& Noels A. 1993, in Origin and Evolution of the Elements, ed.
N. Prantzos, E. Vangioni-Flam, \& M. Casse (Cambridge: Cambridge Univ. Press), 15
\bibitem [\protect\citeauthoryear{Grevesse \& Sauval}{1998}] {gs98} Grevesse N., \& Sauval A.J., 1998, Space Sci. Rev., 85, 161
\bibitem [\protect\citeauthoryear{Hinze}{1975}] {hin75} Hinze J.O., 1975, Turbulence, 2nd edn. McGraw-Hill, New York
\bibitem [\protect\citeauthoryear{Hobbs \& Pilachowski}{1986}] {hp86} Hobbs L.M., \& Pilachowski C., 1986, ApJ, 309, L17
\bibitem [\protect\citeauthoryear{Hobbs et al.}{1989}] {hob89} Hobbs L. M., Iben I., \& Pilachowsky C., 1989, ApJ, 347, 817
\bibitem [\protect\citeauthoryear{Iglesias \& Rogers}{1996}] {igl96} Iglesias C.A., \& Rogers F.J., 1996, ApJ, 496, L121
\bibitem [\protect\citeauthoryear{Jeffries}{1997}] {je97} Jeffries R.D., 1997, MNRAS, 292, 177
\bibitem [\protect\citeauthoryear{Jeffries et al.}{2002}] {je02} Jeffries R.D., Totten E.J., Harmer S., \& Deliyannis C.P., 2002, MNRAS, 336, 1109
\bibitem [\protect\citeauthoryear{Jones et al.}{1999}] {jo99} Jones B.F., Fischer D., \& Soderblom D.R., 1999, AJ, 117, 330
\bibitem [\protect\citeauthoryear{King et al.}{1997}] {ki97} King R.J. et al., 1997, AJ, 113, 1871
\bibitem [\protect\citeauthoryear{King et al.}{2010}] {ki10} King R.J., Schuler S.C., Hobbs L.M., \& Pinsonneault M.H., 2010, ApJ, 710, 1610
\bibitem [\protect\citeauthoryear{Li \& Yang}{2007}] {li07} Li Y.,\& Yang J. Y., 2007, MNRAS, 375, 388
\bibitem [\protect\citeauthoryear{Lubin et al.}{2010}] {lt10} Lubin D., Tytler D., \& Kirkman D., 2010, ApJ, 716, 766
\bibitem [\protect\citeauthoryear{Meakin \& Arnett}{2007}] {mk07} Meakin C.A. \& Arnett D., 2007, ApJ, 667, 448
\bibitem [\protect\citeauthoryear{Meakin \& Arnett}{2010}] {mk10} Meakin C.A.,\& Arnett W.D., 2010, ApSS, 328, 221
\bibitem [\protect\citeauthoryear{Mel\`{e}ndez \& Ram\'{\i}rez}{2007}] {mr07} Mel\`{e}ndez J., \& Ram\'{\i}rez I., 2007, ApJL, 669, 89
\bibitem [\protect\citeauthoryear{Michaud \& Charbonneau}{1991}] {mi91} Michaud G.,\& Charbonneau, P., 1991, Space Science Reviews , 57, 1
\bibitem [\protect\citeauthoryear{Michaud}{1986}] {mi86} Michaud G. 1986, ApJ, 302, 650
\bibitem [\protect\citeauthoryear{Montalb\'{a}n}{1994}] {mon94} Montalb\'{a}n J. 1994, A\&A, 281, 421
\bibitem [\protect\citeauthoryear{Montalb\'{a}n \& Schatzman}{1996}] {mon96} Montalb\'{a}n J., \& Schatzman, E., 1996, A\&A, 351, 347
\bibitem [\protect\citeauthoryear{Montalb\'{a}n \& Schatzman}{2000}] {mon00} Montalb\'{a}n J., \& Schatzman, E., 2000, A\&A, 354, 943
\bibitem [\protect\citeauthoryear{Morales et al.}{2008}] {mo08} Morales J.C., Ribas I., \& Jordi C. 2008, A\&A, 478, 507
\bibitem [\protect\citeauthoryear{Morales et al.}{2010}] {mo10} Morales J.C., Gallardo J., Ribas I., Jordi C., et al., 2010, ApJ, 718, 502
\bibitem [\protect\citeauthoryear{Moss}{1968}] {mo68} Moss D.L. 1968, MNRAS, 141, 165
\bibitem [\protect\citeauthoryear{Paczynski}{1969}] {paz69} Paczynski B., 1969, Acta Astr., 19, 1
\bibitem [\protect\citeauthoryear{Pasquini et al.}{2001}] {pa01} Pasquini L., et al. 2001, A\&A, 374, 1017
\bibitem [\protect\citeauthoryear{Paxton et al.}{2011}] {pa11} Paxton B., Bildsten L., Dotter A., Herwig F., Lesaffre P., \& Timmes F., 2011, ApJS, 192, 3
\bibitem [\protect\citeauthoryear{Perryman et al.}{1998}] {pe98} Perryman M.A.C., Brown A.G.A., Lebreton Y., et al. 1998, A\&A, 331, 81
\bibitem [\protect\citeauthoryear{Piau \& Turck-Chi\`{e}ze}{2002}] {pt02} Piau L., \& Turck-Chi\`{e}ze, S., 2002, ApJ, 566, 419
\bibitem [\protect\citeauthoryear{Pinsonneault et al.}{1990}] {pin90} Pinsonneault M. H., Kawaler S. D., \& Demarque P., 1990, ApJS, 74, 501
\bibitem [\protect\citeauthoryear{Pinsonneault et al.}{1999}] {pin99}Pinsonneault M.H. et al., 1999, ApJ, 527, 180
\bibitem [\protect\citeauthoryear{Pinsonneault}{2010}] {pin10}Pinsonneault M.H., 2010, in IAU Symp. 268, Light Elements in
the Universe, ed. C. Charbonnel et al. (Cambridge: Cambridge University Press), 375
\bibitem [\protect\citeauthoryear{Proffitt \& Michaud}{1989}] {pm89}Proffitt C.R., \& Michaud G., 1989, ApJ, 346, 976
\bibitem [\protect\citeauthoryear{Randich et al.}{1998}] {ra98} Randich S., Mart\'{\i}n E., Garc\'{\i}a L\'{o}pez R.J., \& Pallavicini R. 1998, A\&A, 333, 591
\bibitem [\protect\citeauthoryear{Randich et al.}{2000}] {ra00} Randich S., Pasquini L., \& Pallavicini R., 2000, A\&A, 356, L25
\bibitem [\protect\citeauthoryear{Ribas}{2006}] {ri06} Ribas I., 2006, Ap\&SS, 304, 89
\bibitem [\protect\citeauthoryear{Richard et al.}{2002}] {ri02} Richard O., Michaud G., Richer J., Turck-Chi\`{e}ze S., \& VandenBerg D.A.,  2002, ApJ, 568, 979
\bibitem [\protect\citeauthoryear{Richard et al.}{2005}] {ri05} Richard O., Michaud G., \& Richer J., 2005, ApJ, 619, 538
\bibitem [\protect\citeauthoryear{Rogers et al.}{1996}] {rog96} Rogers F.J., Swenson F.J., \& Iglesias C.A., 1996, ApJ, 456, 902
\bibitem [\protect\citeauthoryear{Schlattl \& Weiss}{1999}] {sch99} Schlattl H., \& Weiss A., 1999, A\&A, 347, 272
\bibitem [\protect\citeauthoryear{Schramm et al.}{1990}] {sch90}  Schramm D. N., Steigman G., \& Dearborn D. S. P., 1990, ApJ, 359, L55
\bibitem [\protect\citeauthoryear{Sestito \& Randich}{2005}] {sr05} Sestito P., \& Randich S., 2005, A\&A, 442, 615
\bibitem [\protect\citeauthoryear{Sestito et al.}{2004}] {se04} Sestito P., Randich S., \& Pallavicini R., 2004, A\&A, 426, 809
\bibitem [\protect\citeauthoryear{Singh et al.}{1995}] {si95} Singh H.P., Roxburgh I.W., Chan K.L., 1995, A\&A, 295, 703
\bibitem [\protect\citeauthoryear{Soderblom et al.}{1993a}] {so93} Soderblom D.R., Fedele S.B., Jones B.F., Stauffer J.R., \& Prosser C.F., 1993, AJ, 106, 1080
\bibitem [\protect\citeauthoryear{Soderblom et al.}{1993b}] {so93b} Soderblom D.R., Jones B.F., Balachandran S., et al., 1993b, AJ, 106, 1059
\bibitem [\protect\citeauthoryear{Straus et al.}{1976}] {str76} Straus J. M., Blake J. B., \& Schramm D. N., 1976, ApJ, 204, 481
\bibitem [\protect\citeauthoryear{Swenson \& Faulkner}{1992}] {swe92}  Swenson F. J., \& Faulkner J., 1992, ApJ, 395, 654
\bibitem [\protect\citeauthoryear{Taylor}{2007}] {tay07} Taylor, B.J., 2007, AJ, 133, 370
\bibitem [\protect\citeauthoryear{Torres \& Ribas}{2002}] {to02} Torres G., \& Ribas I., 2002, ApJ, 567, 1140
\bibitem [\protect\citeauthoryear{Torres et al.}{2006}] {to06} Torres G., Lacy C. H., Marschall L.A., \& et al., 2006, ApJ, 640, 1018
\bibitem [\protect\citeauthoryear{Ventura et al.}{1998a}] {ve98} Ventura P., Zeppieri A., Mazzitelli I., \& D¡¯Antona F. 1998a, A\&A, 334, 953
\bibitem [\protect\citeauthoryear{Ventura et al.}{1998b}] {ve98b} Ventura P., Zeppieri A., Mazzitelli I., \& D¡¯Antona F. 1998b, A\&A, 331, 1011
\bibitem [\protect\citeauthoryear{Xiong}{1985}] {xio85} Xiong D.R., 1985, A\&A, 150, 133
\bibitem [\protect\citeauthoryear{Xiong}{1989}] {xio89} Xiong D.R., 1989, A\&A, 213, 176
\bibitem [\protect\citeauthoryear{Xiong \& Deng}{2001}] {xio01} Xiong D.R.,\& Deng L., 2001, MNRAS, 327, 1137
\bibitem [\protect\citeauthoryear{Xiong \& Deng}{2002}] {xio02} Xiong D.R.,\& Deng L., 2002, MNRAS, 336, 511
\bibitem [\protect\citeauthoryear{Xiong \& Deng}{2005}] {xio05} Xiong D.R.,\& Deng L., 2005, ApJ, 622, 620
\bibitem [\protect\citeauthoryear{Xiong \& Deng}{2009}] {xio09} Xiong D.R.,\& Deng L., 2009, MNRAS, 395, 2013
\bibitem [\protect\citeauthoryear{Yang \& Li}{2007}] {yan07} Yang J.Y.,\& Li Y., 2007, MNRAS, 375, 403
\bibitem [\protect\citeauthoryear{Yang \& Bi}{2007}] {yang07}  Yang W.M. \& Bi S.L., 2007, ApJ, 658, L67
\bibitem [\protect\citeauthoryear{Zhang \& Li}{2009}] {zha09} Zhang Q.S., \& Li Y., 2009, RAA, 9, 585
\bibitem [\protect\citeauthoryear{Zhang \& Li}{2012a}] {zha12a} Zhang Q.S., \& Li Y., 2012a, ApJ, 746, 50
\bibitem [\protect\citeauthoryear{Zhang \& Li}{2012b}] {zha12b} Zhang Q.S., \& Li Y., 2012b, ApJ, 750, 11
\end{thebibliography}
\end{document}